\documentclass{jkas}
\usepackage{multirow}
\usepackage{amsmath}
\usepackage{algorithm}
\usepackage{algorithmic}
\usepackage{graphicx}
\usepackage[colorlinks=true,citecolor=blue]{hyperref}

\def\year{2018} 
\def\month{July} 
\def\volume{00} 
\def\issue{0} 
\def\beginpage{1} 
\def\endpage{99} 
\def\received{---} 
\def\accepted{---} 
\date{Received \received; accepted \accepted}

\title{
VLBI Network SIMulator: An Integrated Simulation Tool for Radio Astronomers
}

\author[1]{Zhen Zhao}
\author[1,2]{Tao An}
\author[1]{Baoqiang Lao}

\affil[1]{Shanghai Astronomical Observatory, Chinese Academy of Sciences, 200030 Shanghai, China;}
\affil[2]{Key Laboratory of Radio Astronomy, Chinese Academy of Sciences, 210008 Nanjing, China;}

\def\corrauthor{
T. An, \email{antao@shao.ac.cn}
}

\def\runningauthor{
Zhao, An, \& Lao
}

\def\runningtitle{
VLBI Network SIMulator
}

\def\keywords{
journals: simulation: VLBI, Interferometry, image simulation
}

\def\abstracttext{
In this paper we introduce a software package, the Very Long Baseline Interferometry Network SIMulator (VNSIM), which provides an integrated platform assisting radio astronomers to design the Very Long Baseline Interferometry (VLBI) experiments and evaluate the network performance with a user-friendly interface. Though VNSIM is motivated to be designed for the East Asia VLBI Network, it can also be expandable to other VLBI networks and generic interferometers. The software package not only integrates the functionality of plotting $(u,v)$ coverage, scheduling the observation, and displaying the dirty and CLEAN images, but also further extends to add new features such as the sensitivity calculation of a certain network and multiple-satellite space VLBI simulations which are useful for future space VLBI mission. In addition, VNSIM provides flexible interactions on both command line and graphical user interface and offers friendly support for log report and database management. VNSIM also supports multiprocessing accelerations, enabling to handle large survey data. To facilitate the future development and update, each simulation function is encapsulated in different Python module allowing for independently invoking and testing. In order to verify the performance of VNSIM, we have carried out various simulations and compared the results with other simulation tools. All tests show good consistency. }

\begin{document}
\jkashead 


\section{Introduction}

East Asia Very Long Baseline Interferometry (VLBI) Network \citep[EAVN,][]{bgEavn}, consisting of 21 radio telescopes from China, Japan and South Korea, has been astronomically operational since mid -2018. The diverse sub-array configurations and frequency setups of the EAVN make it cover a wide range of research areas including astronomical masers (e.g. hydroxyl, methanol, water and SiO masers in Galactic objects and extragalactic megamasers as well), jets of active galactic nuclei, pulsars and transients (e.g. supernovae, gamma-ray bursts), space exploration and tracking, astrometry and geodesy. The EAVN is expected to promote the regional collaborations in East Asia. Such academic collaborations will offer great opportunities to expand the discovery fields in astronomy and space science, and form a successful model of international academic collaborations with a sustainable operation scheme.

Along with the formal operation of the EAVN, a set of user assistance software, having functions of evaluating the network performance, would be helpful for users to prepare proposals and necessary for expanding the user community. The diversity and versatility of the EAVN configurations demand such a tool to be sufficiently flexible and expandable.

Currently, there are a number of auxiliary tools used in the VLBI community for different simulation purposes. SCHED\footnote{http://www.aoc.nrao.edu/software/sched/} designed by the National Radio Astronomy Observatory (NRAO) of the US, is commonly used for scheduling VLBI observations. It supports plotting the $(u, v)$  coverage of a given array and time range, the corresponding dirty beam, and displaying the telescope elevation angle change with time which assists scientists to choose proper telescopes. SCHED can also show multiple source scans in time sequence, which facilitates the arrangement of time blocks to optimize the $(u, v)$ coverage within the allocated time period. Since SCHED is configured through importing a pre-defined 'key' file, any parameter change or setup adjustment requires restarting the program. Difmap \citep{softDifmap} is largely used to analyze VLBI data, as a part of the Caltech VLBI Analysis Programs \citep{bgVLBI1, bgVLBI2}. It is characteristic of interactive operations with editing, hybrid imaging, self-calibration and automated pipeline capacities.  Difmap provides both scripting language and commands for operations. Although these software packages have been and are still widely used in VLBI community, the architecture and algorithms of SCHED and Difmap are slightly outdated. The aperture synthesis simulator (APSYNSIM) \citep{softApsynsim} is a Python-based software package, providing an interactive tool to visualize the aperture synthesis and perform educational level simulations which are very useful for non-radio astronomers.

Besides these auxiliary tools for ground-based VLBI networks, some other software packages were developed with additional functions specially to adapt to space VLBI. These include: the Space VLBI Assistance Software (SPAS) developed by the Satellite Geodetic Observatory of the Institute of Geodesy \citep{softSpas} and {\it Fakesat} \citep{bgfakesat} for VLBI Space Observatory Programme  \citep{bgVsop} proposal preparation; Astronomical Radio Interferometer Simulator (ARIS) developed by the Japan Aerospace Exploration Agency \citep{softAris1, softAris2} for Japan's second-generation space VLBI project VSOP2, which adds some new  functions to assess the impacts of a variety of error sources on the image quality; the FakeRat software package developed by Astro Space Center of the Lebedev Physical Institute \citep{softFaset} and used for the Russia RadioAstron  \citep{bgRadio} space VLBI mission.
Since these works are dedicate for specific space missions, thus their general versatility is relatively limited.
Shanghai Astronomical Observatory of China once proposed the space millimeter-wavelength VLBI array (SMVA) programme which for the first time involves two space satellites onboard 10-m radio telescopes operated at the highest frequency of 43GHz  \citep{bgHong}.
In order to support this SMVA, a simulation software \citep{softShaouv} was designed to provide fundamental space-ground and space-space VLBI $(u,v)$ coverage simulations.

The major functions of these existing software packages which are adapted for specific applications have been discussed and compared in \citet{softShaouv}. For these tools to work for a new VLBI network such as the EAVN, major modifications have to be made. Moreover, a module-independent, highly scalable, flexible, and user-friendly solution, combining the advantages of each individual software and data-modifying flexibility, is highly desirable. For this purpose, we developed a new software package, VLBI Network SIMulator (VNSIM), which integrates most commonly-used simulation programs and offers assistance for radio astronomers. It is a cross-platform Python-based software package with high scalability and reusability. Each function has been separately designed and independently implemented for the sake of future extension.
In addition to common simulation functions,  VNSIM also augments some extended functions such as displaying all-year-round $(u,v)$  plots to track the space VLBI $(u,v)$ coverage changes due to the satellite orbit precession and presenting $(u,v)$ plots of multiple sources along with their dirty maps so as to evaluate the imaging performance.
The data of stations and sources are dynamically managed by the SQLite database, and the parameter configuration can either be saved and loaded directly or be set up through the interactive graphical interfaces.
The data processing with multiple processing accelerations is also adopted, significantly enhancing the execution performance of large survey data.
The simulations of EAVN are performed in a straightforward way.
Although VNSIM is initially designed for EAVN, it is naturally adapted to other VLBI networks or other generic interferometers.

The remaining part of this paper is organized as follows. Section 2 describes the overall designing concept of VNSIM. Details of the main functions are presented in Section 3. In Section 4, we demonstrate  some examples of the major functionality of VNSIM and also compare the experimental results with other tools. A summary is given in Section 5.

\section{The VNSIM}

\begin{figure*}
\centering
\includegraphics[width=\textwidth]{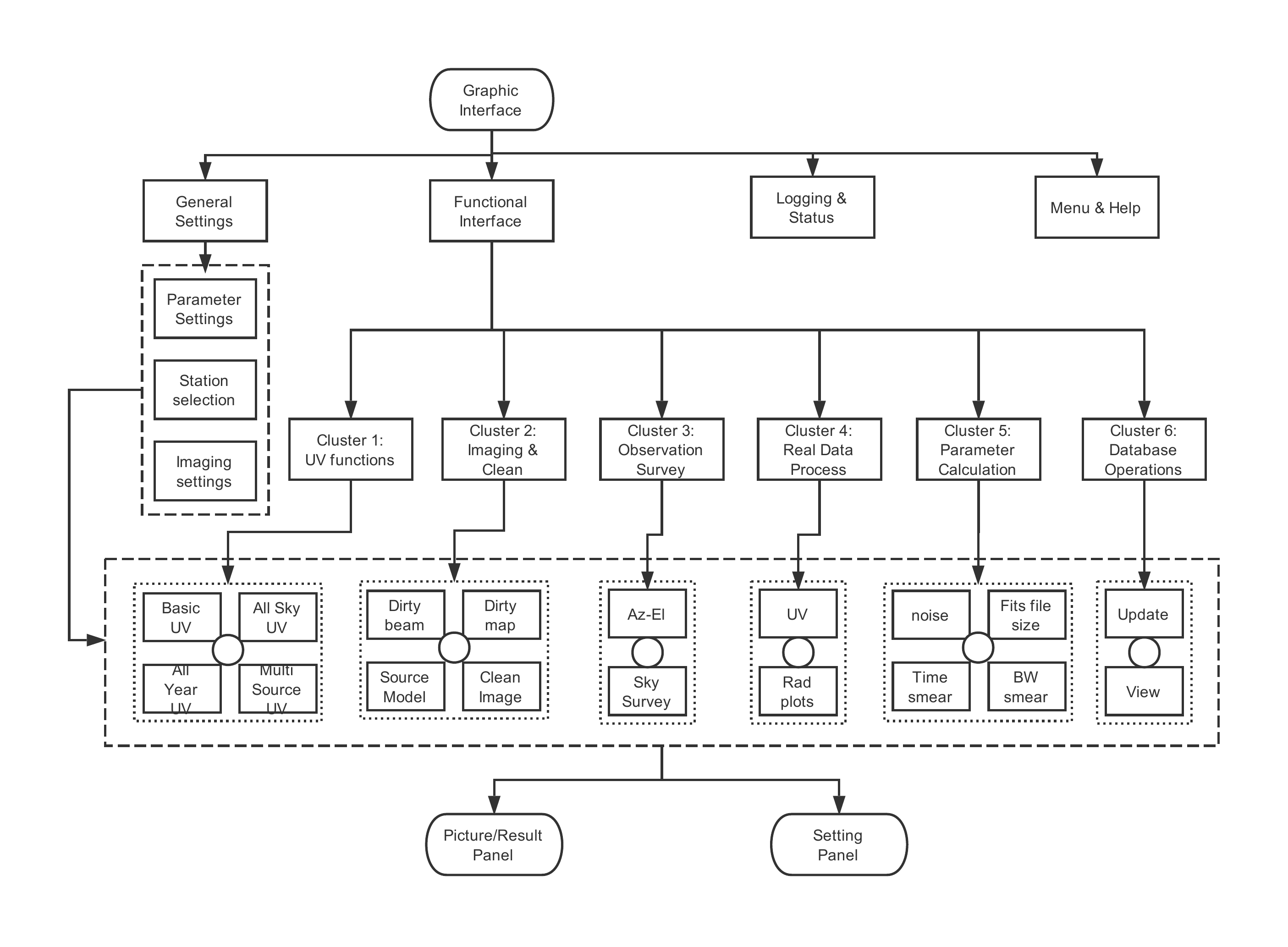}
\caption{The software framework of VNSIM\label{fig:framework}. This diagram shows the example of the graphical user interface. Detailed description of the main functions are given in Section 3. VNSIM currently comprises six function clusters including 14 sub-functions in total.}
\end{figure*}

The VNSIM is programmed in Python language and can be implemented on all common platforms, Windows, Linux and Mac OS. The source code is available on Github\footnote{https://github.com/ZhenZHAO/EAVNSIM}. In this section we describe the overall design and framework of the VNSIM.

\subsection{Overall Design}
To fully match the requirements of simulating EAVN tasks, we mainly propose the following design considerations:
\begin{itemize}
\item Highly scalable and reusable. Each function cluster is independently encapsulated into a single Python file to facilitate the testing and future extensions. In this sense, the design obeys the Model View Controller (MVC) architectural pattern. We divide the development into three interconnected parts, {\it i.e.}, an SQLite database for data management, a graphical interface designed by Tkinter for user interactions, and functional clusters for parameter calculations.

\item Dynamic database management. The information of VLBI stations and astronomical sources is stored and managed by an SQLite database, which is implemented through a graphical user interface (GUI), enabling to add, delete and modify existing data records in a convenient way. It also allows for appending self-defined fake stations for the simulation purpose. The related database currently consists of four tables: source table, VLBI station table, satellite table and telemetry station table.
The first two are used for ground- and space-based VLBI network simulation, and the last two are only needed for space VLBI mission.
The data can either be inserted through the GUI record by record or be directly added as a whole through importing external files in .txt or .csv formats.

\item User-friendly interface. Both graphic and command line interfaces are supported in VNSIM for different-level users with different usage habits. In either mode, the parameter configurations can be saved and loaded conveniently, and all the resulting images can be saved in a variety of image formats. Besides, indicative logging information and dialogues are designed to help users to track the running status of VNSIM.

\item Running performance. Unlike SCHED, VNSIM is compatible with generic interferometers, such as Very Large Array (VLA) and the Square Kilometre Array (SKA). VNSIM contains some new advanced functions which are suitable for $(u,v)$ coverage simulations of large scale sky survey of these many-element interferometers, for long-term evaluation of the $(u,v)$ coverage of main targets in the space VLBI. Undoubtedly these functions involve huge calculations and produce large-size images, typically consuming considerable computing resources. To address these problems, we consider the compute-intensive accelerations via multiple  processing. In addition, multiple threads are created to maintain the graphical interface and functional calculation separately.

\item Functional designs. To deliver a complete solution, VNSIM aims to integrate the most common simulation functions, including the $(u,v)$ coverage plots of a single source and multiple sources, all-sky $(u,v)$ plots of a large sample, and all-year-round $(u,v)$ coverage changes of a single source with the precession of satellite orbits; image simulations of dirty beam, dirty map and CLEAN map; scheduling setups showing azimuth and elevation changes of sources in VLBI stations with observing time; the visibility amplitude of the simulated data as a function of the projected $(u,v)$ distance. Besides, VNSIM also enables us to estimate the VLBI network performance, such as the sensitivity estimate with the given time range and data rate.

\end{itemize}

\subsection{Software Framework}
\begin{figure}
\centering
\includegraphics[width=\columnwidth]{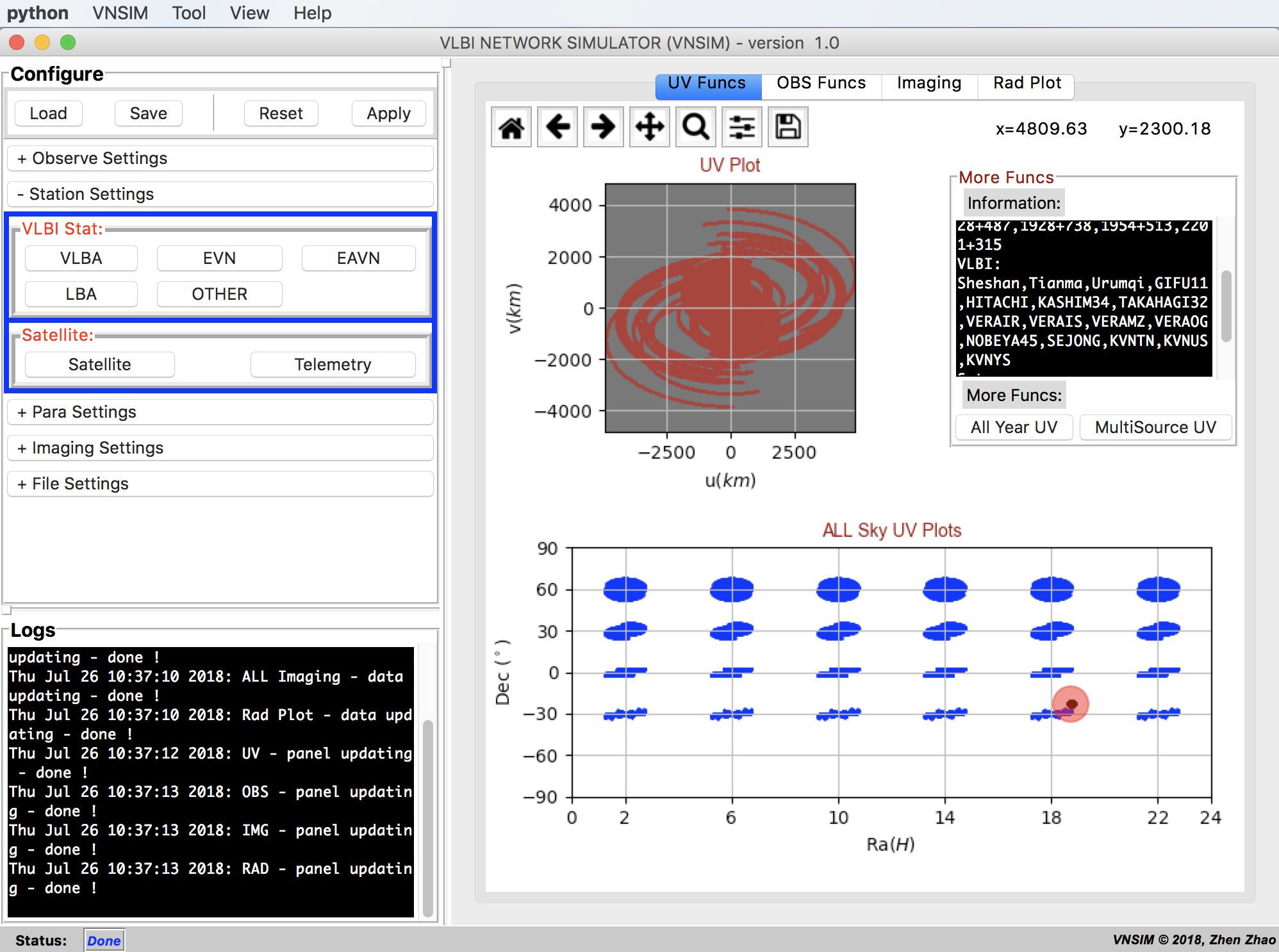}
\caption{The main window of VNSIM\label{fig:main_gui}. It consists of three functions zones: configuration zone, log zone, display zone. The plot in the top-right panel shows the EAVN  $(u,v)$ coverage of 0134+329. More $(u,v)$ coverages can be invoked by changing the source in the configuration panel (\'Observe Settings\'). The bottom-right panel shows the simulations of the EAVN $(u,v)$ coverages of sources at different RA and Dec coordinates. The red circle marks the position of the Sun in the plane of the sky at the observation epoch.}
\end{figure}

As shown in Figure~\ref{fig:framework}, the framework of VNSIM mainly consists of six function clusters, parameter configuration and other common GUI components such as status bar, menu bar, and logging information. The six clusters answer for four fundamental simulation functions, including $(u,v)$ coverage plotting, imaging, observation survey and simulation of visibility data (clusters 1-4), and two additional functions with parameter evaluation and database management (clusters 5-6). As mentioned before, the programming of these function clusters are mutually independent, so each of them can be directly invoked with appropriate configurations and run separately.

As per Figure~\ref{fig:main_gui}, the main area of the VNSIM GUI is divided into three parts: result panel, configuration panel and logging panel. Four main simulation functions locate at different tabs on the result panel, while two additional functions can be accessed through the tool menu. The configure panel comprises of five different kinds of configuring items, and only one of them is visible by default with others folded up. The logging panel shows the currently running information in real-time to help track the VNSIM status. The whole interface is resizable and can be adjusted according to the display window on different computers.

\section{Major Functions}
VNSIM currently comprises six function clusters including 14 sub-functions in total. Those functions are described in detail below.

\subsection{Simulation Functions}
\subsubsection{$(u,v)$ coverage plotting}
The projection of the separation between any two antennas in a VLBI network (so-called baseline, $B$) perpendicular to the direction of the obesrved radio source can be decomposed into east--west and north--south components, represented by $u$ and $v$, respectively. Given the position $(x_\lambda, y_\lambda, z_\lambda)$ of baseline vector $B_\lambda$ in $(X, Y, Z)$ coordinate system, the components $(u,v,w)$ are obtained by \citep{uvCoverage},
\begin{equation*}
\begin{pmatrix}
u \\
v \\
w \\
\end{pmatrix}
\!=\!
\begin{pmatrix}
\sin H_s 					& \cos H_s						& 0 \\
-\sin \delta_s \cos H_s		& \sin \delta_s \sin H_s		& \cos \delta_s \\
\cos \delta_s \cos H_s		& - \cos \delta_s \sin H_s		& \sin \delta_s
\end{pmatrix}\!
\begin{pmatrix}\!
x_\lambda \\
y_\lambda \\
z_\lambda
\end{pmatrix},
\end{equation*}
where $H_s$ and $\delta_s$ are the hour angle and declination of the source position, and $(x_\lambda, y_\lambda, z_\lambda)$ are in unit of the observing wavelength. Each baseline in continuous observations of a radio source will project a portion of an arc of an ellipse on the $(u,v)$ plane, while creates a single curve line in the $(u,v)$ coverage. Multiple stations of VLBI networks form a set of baselines. They together create the usually seen plot in the $(u,v)$ spacing.

In addition to the above simplest $(u,v)$ coverage involving a single source in a certain time period, three additional functions of $(u,v)$ coverage are implemented in VNSIM: (i) all-year-round $(u,v)$ plotting can create twelve $(u,v)$ coverage plots generated through simulating the first-day observation of each month, which is designed to observe the coverage evolution of space VLBI due to the satellite orbit precession; (ii) by evenly dividing the whole sky into $5 \times 6$ blocks, all-sky $(u,v)$ plotting function roughly shows the survey ability of selected station combinations at a given observing time. This is useful for scheduling large-sample survey. In the example shown in Figure 2 bottom-right panel, the $\delta=0^\circ$ source is marginally seen by the EAVN; (iii)  Similarly, the function of multiple-source $(u,v)$ plotting can display the $(u,v)$ coverage of selected multiple sources to assess the performance of a given VLBI network. This is invoked by clicking the button MultiSource UV.

\subsubsection{Imaging}
\begin{figure} [!t]
\centering
\includegraphics[width=0.8\columnwidth]{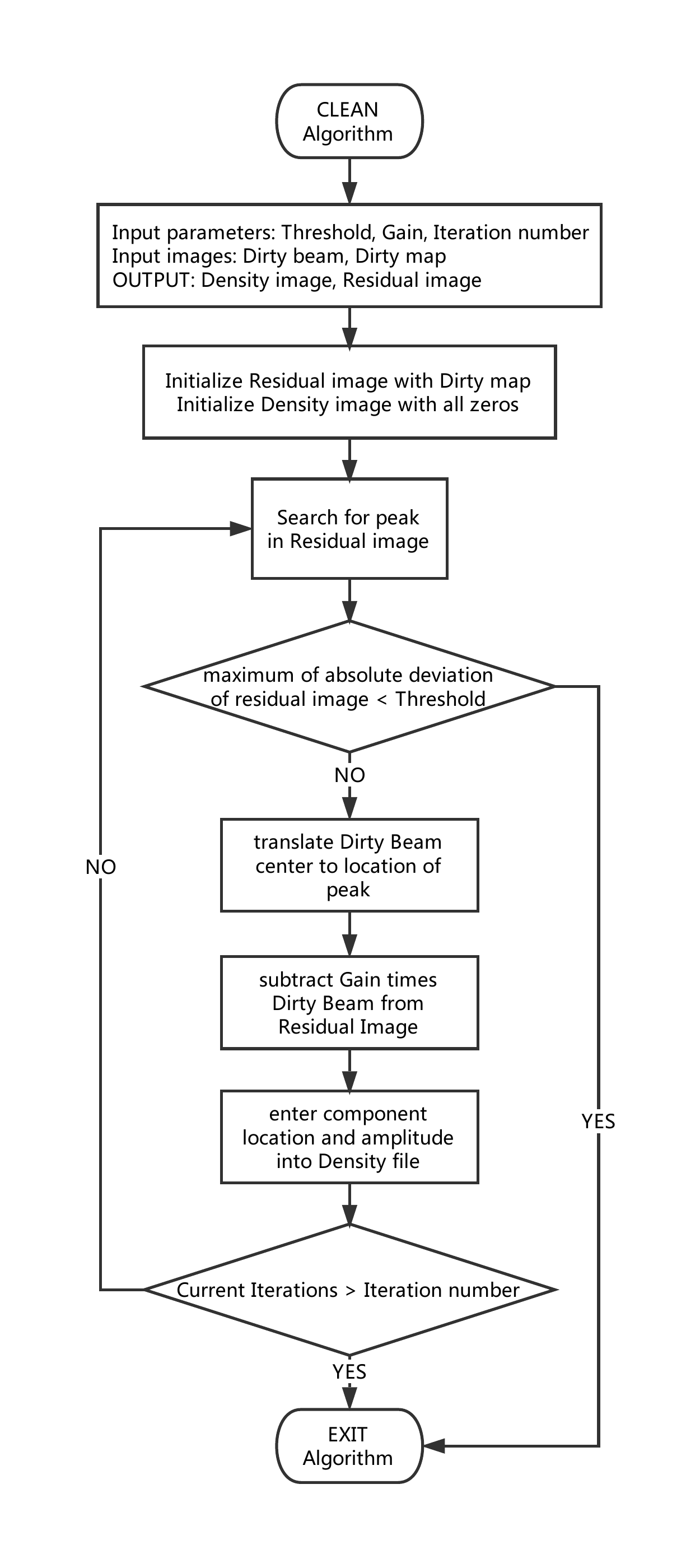}
\caption{Flowchart of the CLEAN loop\label{fig:clean}. }
\end{figure}

The intensity distribution $I_\nu$ of a radio source at observing frequency $\nu$ can be estimated by the two-dimensional Fourier transformation of the spatial coherence function $V_\nu$ \citep{imageThom},
\begin{equation}
I_\nu(l,m) = \iint V_\nu(u,v) e^{2\pi i (ul+vm)} du dv,
\end{equation}
where $(l,m$ is the sky position of the sources.
However, in practice $V_\nu$ is not a continuous function, but is sampled at particular positions on the $(u,v)$ plane. Then, we actually have
\begin{equation}
I_\nu^D(l, m) = \iint V_\nu(u,v) S(u,v) W(u,v) e^{2\pi i (ul+vm)} du dv,
\end{equation}
where $I_\nu^D(l, m)$ is the so-called dirty image; $S(u,v)$ and $W(u,v)$ represent the adopted sampling and weighting functions, respectively. Besides, the observed intensity distribution is often expressed as a convolution of intrinsic dirty image and the synthesized beam or the point spread function of the interferometer $B_\theta$,
\begin{equation}
I_\nu^D(l, m) = I_\nu * B_\theta,
\end{equation}
where $B_\theta$ is given by
\begin{equation}
B_\theta(l,m) = \iint S(u,v) W(u,v) e^{2\pi i (ul+vm)} du dv.
\end{equation}

Given a selected source model (point source model, Gaussian model, etc) and proper observation configurations in simulations, VNSIM can generate the corresponding $(u,v)$ coverage plot, dirty beam $B_\theta$, and dirty image $I_\nu^D$. Referring to \citep{cleanAlgo}, we also integrate the CLEAN algorithm in VNSIM to perform quick inspection of the image quality. As shown in Figure~\ref{fig:clean}, the deconvoluted (CLEAN) algorithm can pass two controlling parameters, {\it i.e.}, the Threshold and Iteration Number which control how to stop the iteration procedure. Eventually, adding the group of CLEAN components back into the final residual image produces the simulated CLEAN image. As mentioned before, this function is useful for evaluating the performance of a new VLBI network, or select in optimized network configuration before site selection. In future, more realistic noise model will be considered in order to make the simulation close to the practical condition.


\subsubsection{Observation}
In the third cluster (Figure~\ref{fig:framework}), observation survey, we implement two kinds of simulations. The first one is to plot the azimuth and elevation angles at selected stations as a function of time within the observation duration. It can clearly manifest the best observing time duration of selected stations for a specific target source. The other is on the sky survey function, which aims to check whether different parts of the sky are visible by a VLBI network or not at a certain epoch. Specifically, it visualizes the number of visible observing stations in different color patch at each divided sky position and also the position of the Sun and Moon during the observing time.

\subsubsection{Real Data Process}
VNSIM is not only a simulation and demonstration tool, but also allows for processing real observing data. For example, by extracting the station position and observed visibility data, VNSIM can draw the plots of $(u,v)$ coverage of the real observing data and the corresponding dirty beam. As a useful demonstration, the figure of \'visibility amplitude -- projected $(u,v)$ distance\' is also provided for astronomers to observe the visibility changes, gain a rough knowledge of the source structure (resolved or unresolved), and diagnose bad data points.

\subsection{Additional Functions}

\begin{figure}[!t]
\centering
{
\begin{minipage}[b]{0.45\textwidth}
\includegraphics[width=\columnwidth]{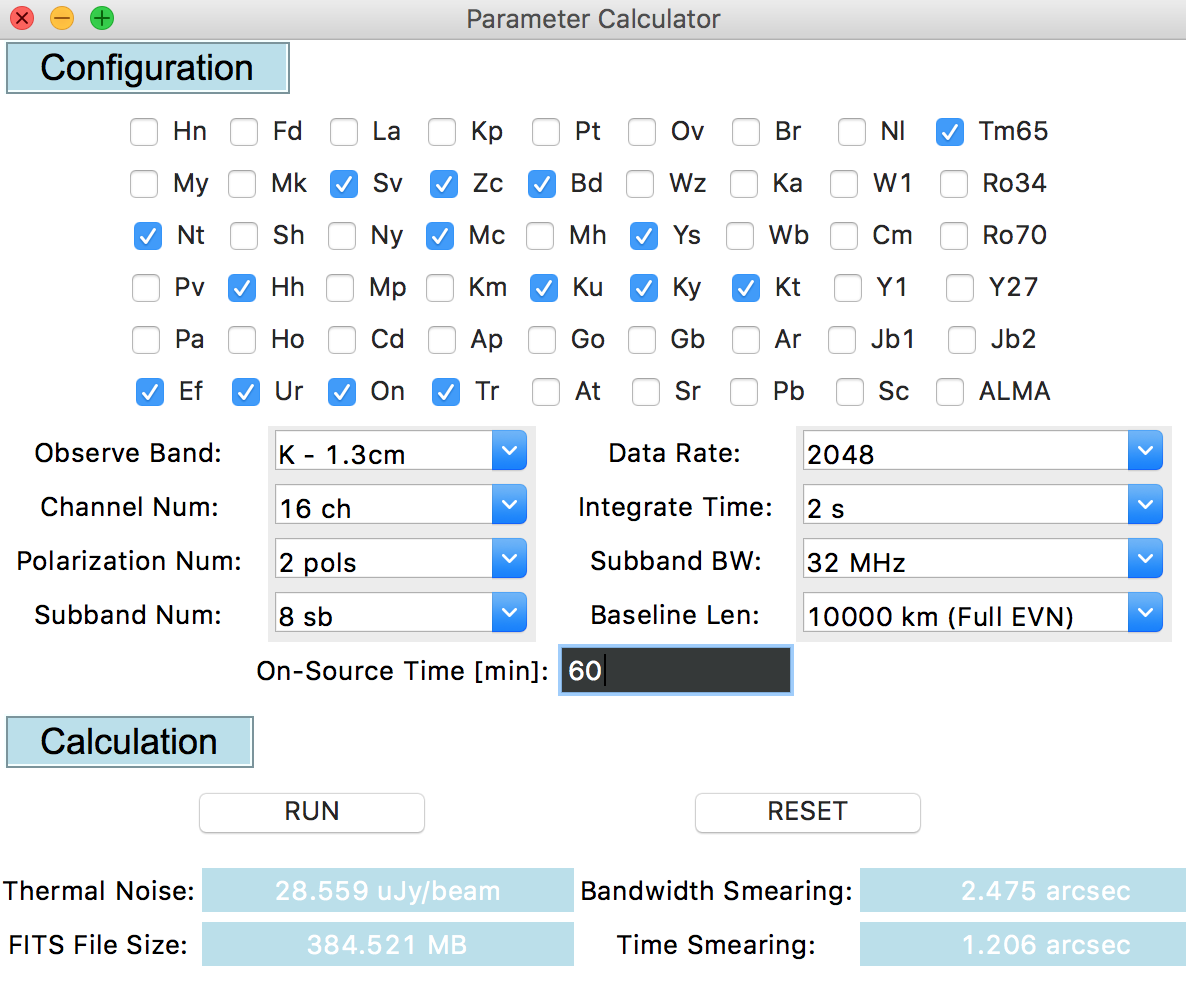}
\end{minipage}
}
{
\begin{minipage}[b]{0.45\textwidth}
\includegraphics[width=\columnwidth]{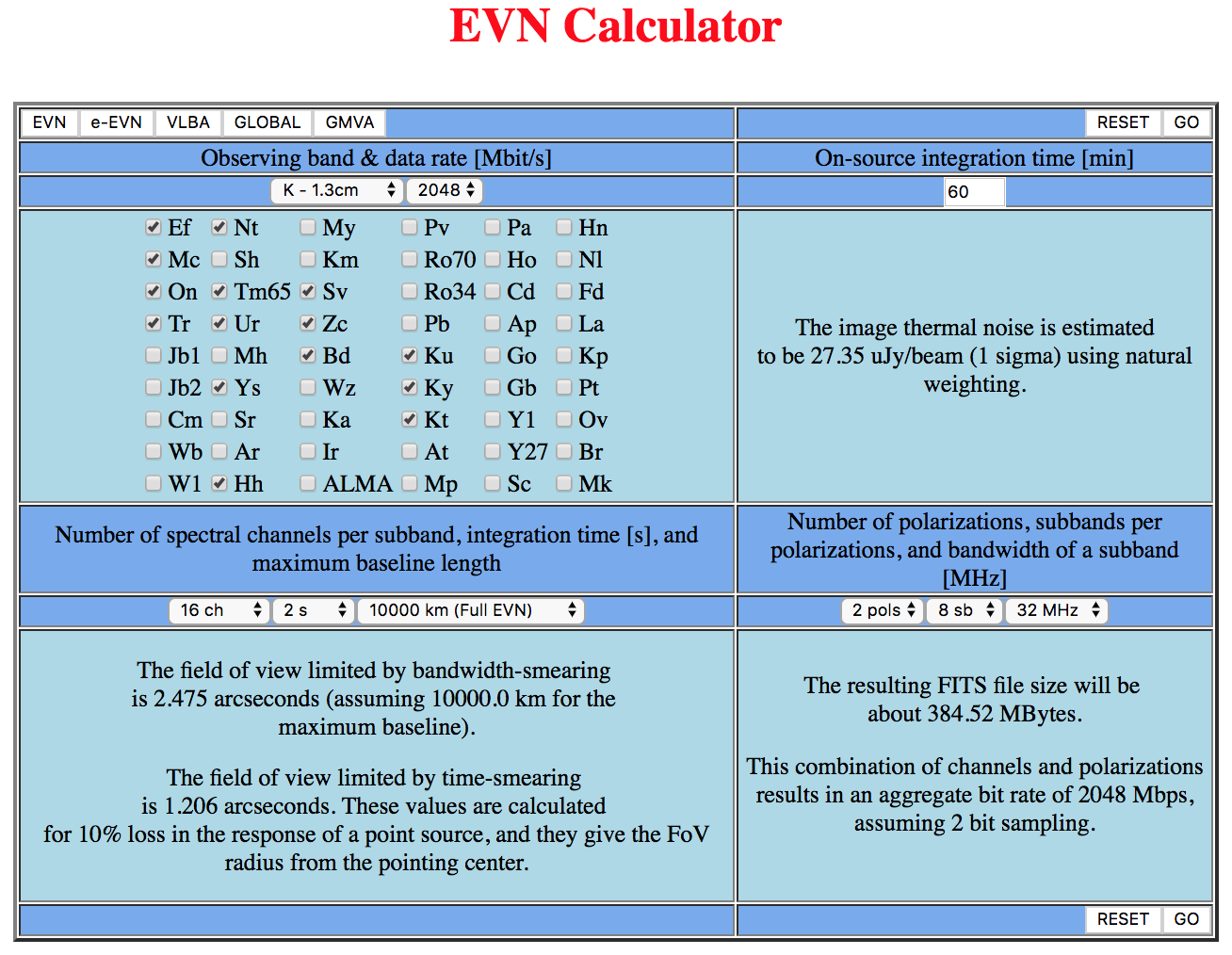}
\end{minipage}
}
\caption{The GUIs of parameter evaluation\label{fig:ui_cal} of VNSIM (top panel)  and the EVN Calculator (bottom panel). The database of VNSIM is dynamically managed, easy to modify.}
\end{figure}

\subsubsection{Parameter Evaluation}

Referring to the EVN Calculator\footnote{http://www.evlbi.org/cgi-bin/EVNcalc}, VNSIM provides the following parameter calculation: the image thermal noise $\Omega$, bandwidth-smearing-limited field of view $F_{bw}$, time-smearing-limited field of view $F_{time}$, and an estimate of the FITS file size $C$.

First of all, the image thermal noise (in unit of Jy beam$^{-1}$) can be calculated by
\begin{equation}
\Omega = \frac{1}{\eta}  \frac{S_e}{\sqrt{\frac{r T_{obs}}{2}}},
\end{equation}
where $r$ and $T_{obs}$ represent the data rate in bit per second (bps) and on-source time in minute, respectively. $\eta$ is a constant. $S_e$ is given by
\begin{equation}
S_e = ( \sqrt{\frac{1}{2} \sum_{i=1}^{N} \sum_{j=1, j\neq i}^N \frac{1}{SEFD_i \times SEFD_j} } )^{-1},
\end{equation}
 where $SEFD_i$ denotes the system equivalent flux density of telescope $i$ in Jy.

The field of view is limited by the bandwidth smearing effect and is expressed by
\begin{equation}
F_{bw} = \kappa_1 \frac{ N_{ch}}{L_{bl} \, BW_{sub}},
\end{equation}
where $N_{ch}$, $L_{bl}$, and $BW_{sub}$ denote the number of channels, maximum baseline length, and the bandwidth of sub-band, respectively. $\kappa_1$ is a constant, which usually equals to $49500$. $F_{bw}$ is in unit of arcsecond.

Similarly, the field of view limited by time-smearing effect is given by
\begin{equation}
F_{time} = \kappa_2 \frac{L_{wave}}{L_{bl}\,T_{int}},
\end{equation}
where $L_{wave}$, $L_{bl}$ and $T_{int}$ represent the wavelength, maximum baseline length, and integration time of correlation, respectively. Constant $\kappa_2$ usually equals to $18560$. Note that these values are calculated by taking into account for 10\% loss in the response of a point source.

To evaluate the size of resulting FITS file, the following equation is applied.
\begin{equation}
C = \sigma \frac{ N_{sta}^2 \, N_{pol} \, N_{sub} \, N_{ch}} {131072\times3600} \frac{T_{obs}}{\,T_{int} }
\end{equation}
where $\sigma$ represents a constant. $N_{sta}$, $N_{pol}$, and $N_{sub}$, $N_{ch}$ denote the number of stations, polarization,  sub-bands and channels per sub-band, respectively. $T_{obs}$ and $T_{int}$ are the total observation time and the integration time of correlation, respectively.

The parameter evaluation interface of the VNSIM is shown in the top panel of Figure 4.
The commonly-used telescopes are included in the default list.
Other telescopes or new ones can be easily added via the database.
The configuration parameters can be selected from the drop-down menu.
The \'on-source time\' text frame accepts manual input.
Clicking the button \'RUN\', the calculated parameters are shown in the bottom lines.
For comparison, we investigated the parameter calculation of the EVN Calculator. As Figure~\ref{fig:ui_cal} shows, typical stations of EVN at K band were selected and the four parameters were calculated and showed on the GUI. The same configurations are applied to the VNSIM calculator too. It is obvious that the calculated results in VNSIM are consistent with those obtained by the EVN Calculator. Note that the slight difference of the image thermal noises derived from the VNSIM and EVN Calculator is due to that SEFDs of some telescopes in these two databases have minor difference.

\subsubsection{Database Managements}
\begin{figure}
\centering
\includegraphics[width=0.9\columnwidth]{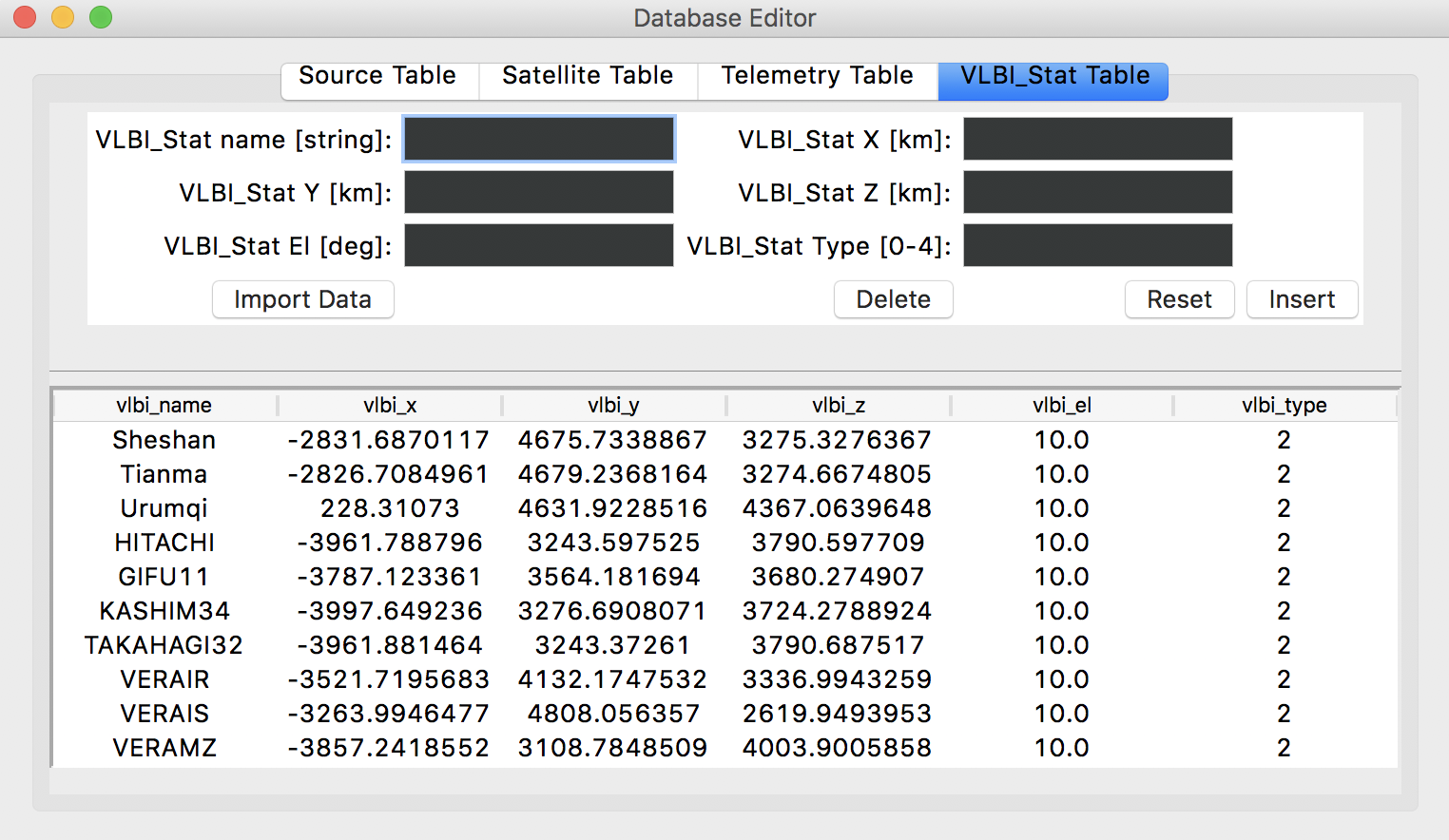}
\caption{The GUI of database editor\label{fig:ui_db} providing a high quality, visual and open source tool to create, delete and edit database files. Users can either edit the records via the GUI, or import a telescope or source table from an external file. }
\end{figure}

Compared with other existing software packages, VNSIM adopts a friendly graphic user interface, allowing for easy access and operation.
We selected SQLite as the SQL database engine.
It is self-contained, highly-reliable, open-source and also full-featured.
The graphic editor of the database management is shown in Figure~\ref{fig:ui_db}. The related database consists of four tables that are responsible for manipulating the corresponding data of the sources, satellites, VLBI stations and telemetry stations, respectively. Taking the VLBI station table as an example, users can delete old records or insert new records easily through the GUI operations. In addition to inserting new data record by record, users can also import multiple records through loading an external file containing well-formatted data.

\section{Example Experiments}

In this section, we take two main functions, $(u,v)$ coverage and source imaging, as examples to demonstrate the operation of VNSIM.

\begin{table}[t!]
\caption{Simulation Parameter Settings\label{tab:para}}
\centering
\begin{tabular}{l|c|p{0.3\columnwidth}c}
\toprule
& Parameters & Settings\\
\midrule
\multirow{3}{2cm}{General}  & Time & 12 h \\
         &  Scan length &  5 min\\
         &  Frequency &  22 GHz  \\ \hline
\multirow{2}{2cm}{Source}  & Main & M87\\
 &  Others$^{\rm a}$  & 0202+319, 0529+483, 1030+415, 1128+385, 1418+546, etc.     \\ \hline

\multirow{2}{2cm}{Station$^{\rm b}$}  & CVN & Tianma, Urumqi, Sheshan \\
 & JVN  & Gifu, Hitachi, Kashima, Takahagi\\
 & KVN  & Sejong, Tamma, Ulsan, Yonsei \\
 & NRO &  Nobeyama\\
 & VERA & Iriki, Ishigakijima, Mizusawa, Ogasawara\\ \hline


\multirow{3}{2cm}{Imaging}  & Iteration & 100 \\
 & Gain & 0.2 \\
 & Threshold & 0.001      \\


\bottomrule
\end{tabular}
\tabnote{
$^{\rm a}$ VNSIM allows for processing multiple sources simultaneously.
In the experimental examples, we take 10 bright and compact AGNs.\\
$^{\rm b}$ the stations are same as in \citet{bgEavn} for comparison.
}
\end{table}

\subsection{$(u,v)$ plotting of a single source}

\begin{figure}[!t]
\centering
{
\begin{minipage}[b]{0.20\textwidth}
\includegraphics[width=\columnwidth]{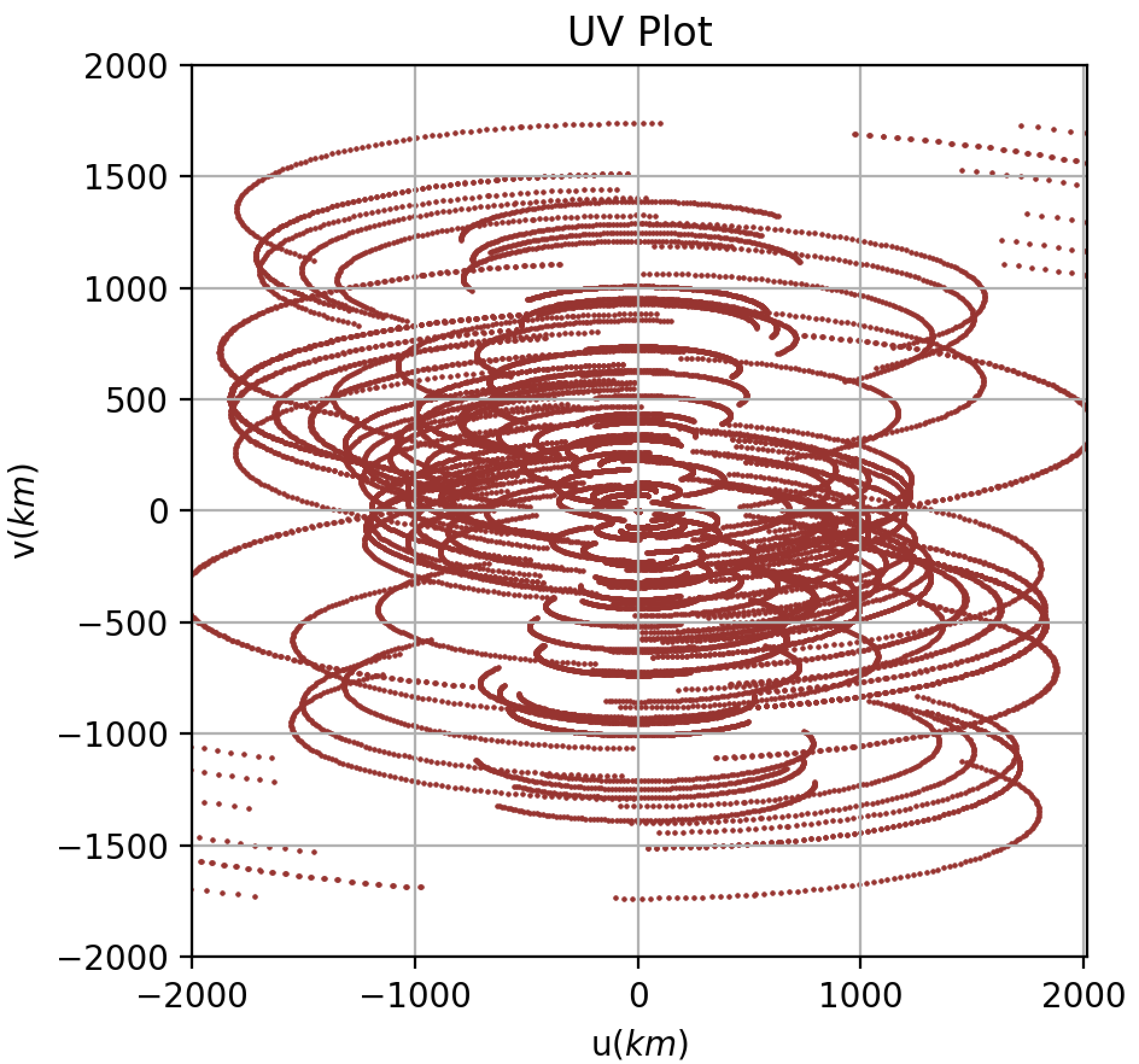}
\end{minipage}
}
{
\begin{minipage}[b]{0.22\textwidth}
\includegraphics[width=\columnwidth]{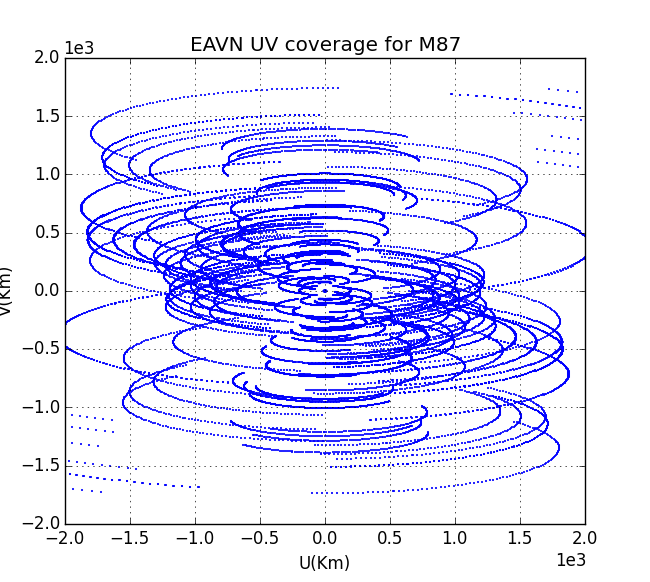}
\end{minipage}
}

\caption{The comparison of $(u,v)$ coverages of EAVN derived from VNSIM ({\it left}, the present paper) and from the literature ({\it right}, \citet{bgEavn}). For the comparison purpose, the inner 2,000 km region of the $(u,v)$ coverage are displayed. They show good consistency.
\label{fig:uv_cmp} }
\end{figure}

$(u,v)$ coverage is a basic visualization of evaluating the VLBI network performance and predicting the image quality.
We first use VNSIM to create the $(u,v)$ coverage plot of a single source, and compare with the existing software.
The radio galaxy M87 is chosen as the target.
As VNSIM supports parallel processing of multiple sources, we also import ten more sources in the source list. The multiple source $(u,v)$ plotting is described in Section 4.2.
The experiment parameters used in this simulation are listed in Table \ref{tab:para}.
For comparison, these parameters are same with those adopted in Figure 2 of \citet{bgEavn}.
The 22 GHz frequency, one of the operational frequency bands in the first EAVN open-use session, is chosen.
A total of sixteen telescopes are included in the simulation.
The 12-hr full-track $(u,v)$ coverages of M87 are shown in Figure~\ref{fig:uv_cmp}.
VNSIM obtained the exactly same $(u,v)$ plot (Figure~\ref{fig:uv_cmp} left panel) with that shown in \citet{bgEavn}.
We note the $(u,v)$ plot in \citet{bgEavn} was made by using the UVSIM software \citep{softShaouv}, whose results have been compared with SCHED.
The consistent results in Figure~\ref{fig:uv_cmp} verify the accuracy of VNSIM and compatibility with other similar software tools.

\subsection{$(u,v)$ plotting of multiple sources}
\begin{figure*}[!t]
\centering
\includegraphics[width=\textwidth]{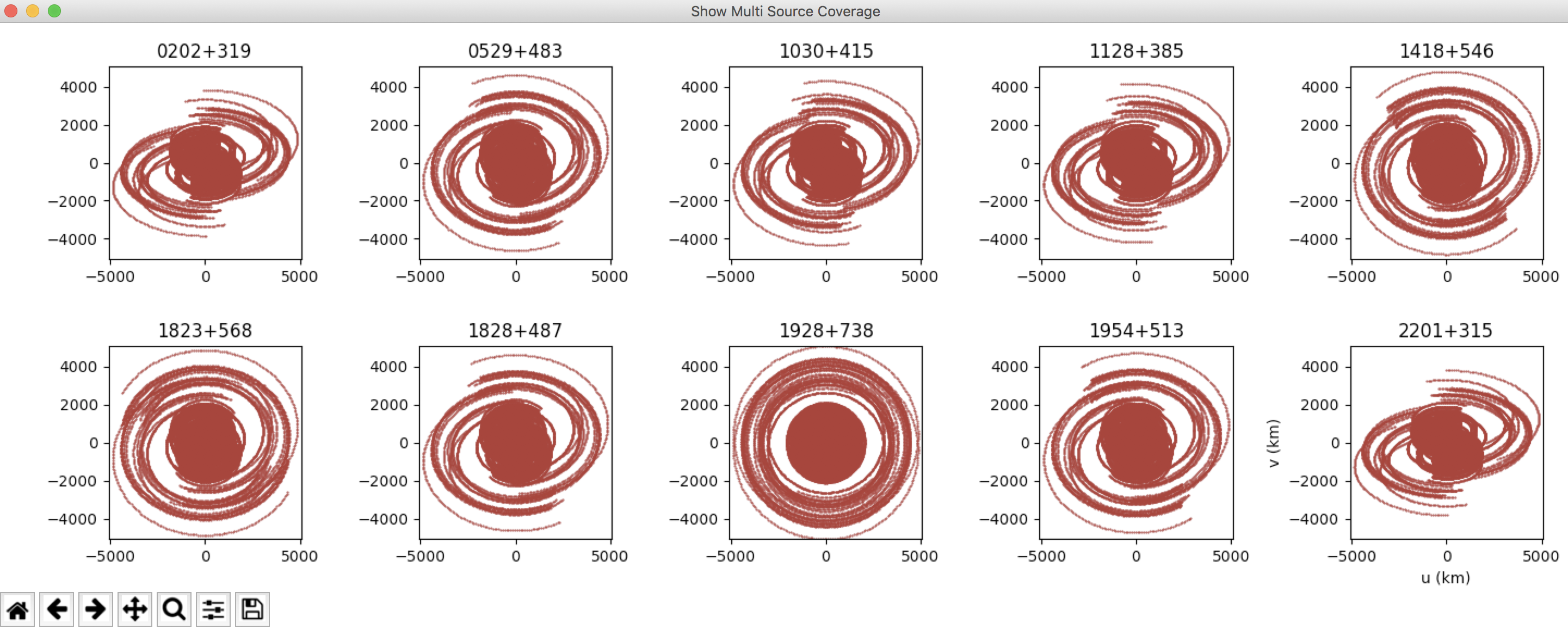}
\caption{An example of $(u,v)$ coverage plotting of multiple sources\label{fig:multiuv}.
This function is useful for large sample surveys.
The present figure shows 10 sources with the same VLBI network configuration in Table 1.
Even more sources are supported by using multiprocess acceleration.
}
\end{figure*}

One of the new practical functions included in VNSIM is the multi-source $(u,v)$ coverage plotting.
It allows for calculating the $(u,v)$ coverages of multiple sources with pre-defined observing configurations through one-shot running and either displays the results in the interface or saves in external files.
Figure~\ref{fig:multiuv} demonstrates an example of the multi-source $(u,v)$ plotting function.
In this experiment, we calculated the $(u,v)$ coverages of ten radio-loud active galactic nuclei selected from \citet{multisource}.
The observing frequency is 43 GHz.
The total observing period of each source in the simulation is 12 h.
The EAVN is used for the VLBI network configuration.
As all these sources are at high declination, they are visible by the EAVN in most time.
In particular, 1823+568 and 1928+738 show almost circular $(u,v)$ coverages.
Moreover, we should mention that the maximal source number is not limited to ten.
Even more sources are allowed and the calculation can be accelerated by multiprocess parallelization.
This functionality is quite useful for large-sample survey.
Besides displaying the result plots in the GUI, the output can also be exported to external '.eps' or other format files.

\subsection{Dirty and CLEAN images}

\begin{table}[t!]
\caption{Four Source Models\label{tab:sources}}
\centering
\begin{tabular}{c|c|c|c|c}
\toprule
 & Relative RA & Relative Dec & Flux Density \\
 & (mas) & (mas) & (Jy) \\
\midrule

1 & 0.0   &0.0  &1.0  \\ \hline
2 & 17.2  &$-$22.7  &0.5 \\ \hline
3 & $-$11.5 &16.5  &1.5 \\ \hline
4 & -18.5 & - 9.5 & 1.5 \\

\bottomrule
\end{tabular}
\end{table}

\begin{figure}[!t]
\centering
\includegraphics[width=\columnwidth]{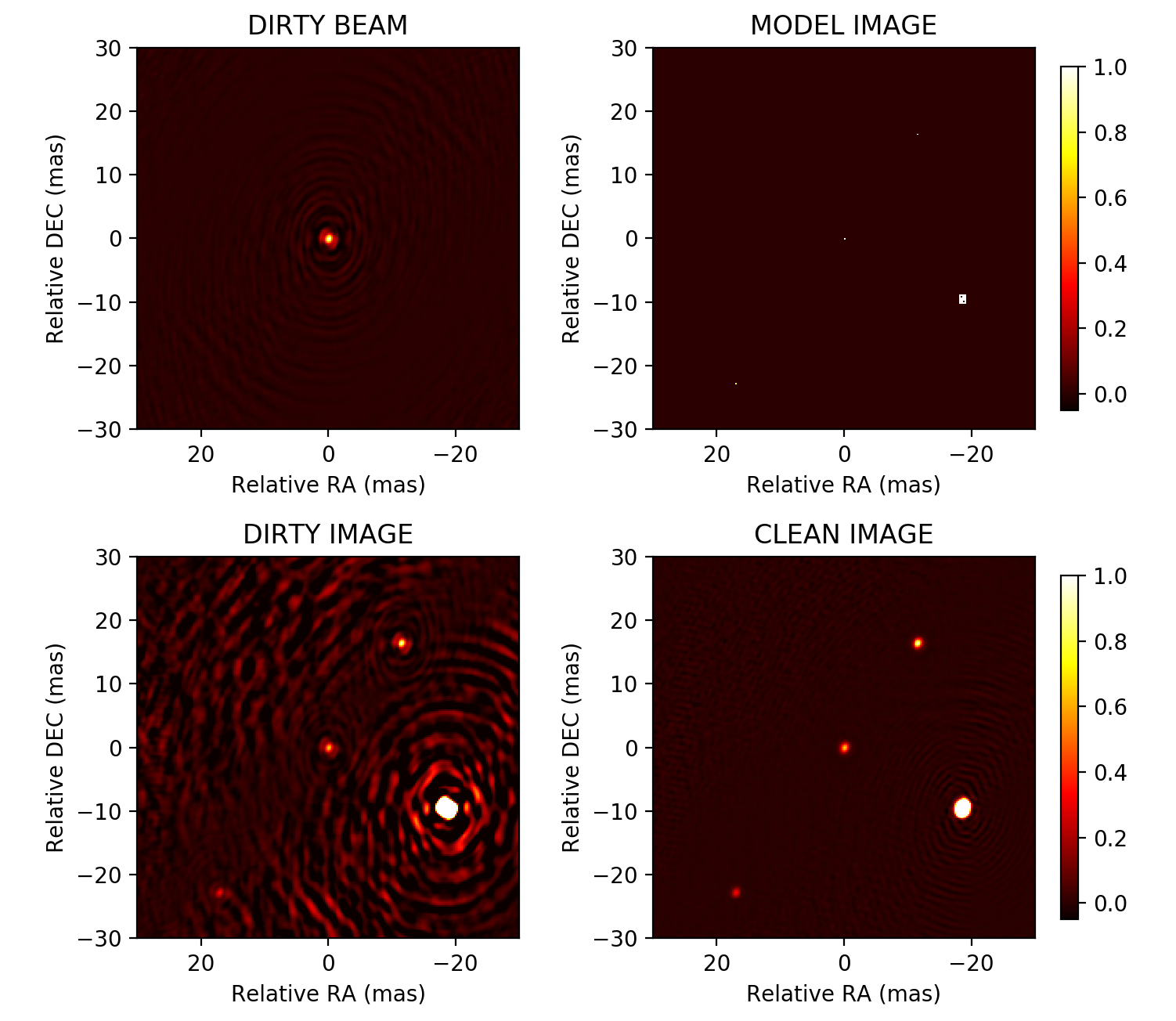}
\caption{Results of the image simulation experiment\label{fig:imaging}. The minor axis of dirty beam size is around $0.653$ mas. Top-right panel shows the distribution of four simulated sources. Bottom-left panel shows the dirty image. The sidelobes around each source are clearly seen. Bottom-right panel shows the deconvoluted image after 100 iterations of CLEANing. The strong sidelobes become less significant, and the dynamic range increases to 237:1.}
\end{figure}

VNSIM is not only a simple display tool of $(u,v)$ coverage, but also a powerful software package enabling image simulation that is visual evaluation of VLBI network imaging performance.
To demonstrate this functionality, we created four fake sources in the experiment whose parameters are listed in Table~\ref{tab:sources}.
As shown in Table~\ref{tab:para}, the same EAVN station and general settings are also applied.
The dirty beam and source models are displayed in the top panel of Figure~\ref{fig:imaging}.
Source \#1 with a flux density of 1 Jy is a point source, locating at the sky position RA$\!=\!$ $03$h$19$m$48.160$s and Dec$\!=\!$ $41^\circ30^\prime42.10^{\prime\prime}$. 
Sources \#2 and \#3 are also point sources and locate at the southeast and northwest directions of Source \#1. Their corresponding flux densities are 0.5  and 1.5 Jy, respectively.
Source \#4 is an extended component, with a size of $0.5 \times \theta_{\rm beam}^{\min}$ and a total flux density of 1.5 Jy, locating at the southwest of the image center.
The bottom-left panel in Figure~\ref{fig:imaging}  depicts the generated dirty image. The ripples of sidelobes of Source \#4 are prominent. 
Sidelobes of other point sources are also visible. 
The bottom-right panel shows the deconvolution image after 100 iterations of CLEANing. The sidelobes have been substantially suppressed, and the background noise becomes smooth. The four sources are distinctive. The dynamic range increases from 11:1 to 237:1. Deeper CLEANing may further decrease the rms noise.

\section{Summary}
In this paper, we have introduced an auxiliary tool aiding VLBI network simulations, named as VNSIM. The motivation is to provide an integrated software package to help radio astronomers to make observation schedule and to gain a preliminary evaluation of the interferometer performance.
Compared with the existing simulation tools, VNSIM not only integrates commonly used functions but also supplements new features supporting large surveys containing multiple sources. Another new feature is the space VLBI simulation which supplies valuable guidance to future space VLBI missions. Details of the space VLBI $(u,v)$ coverage simulation will be presented in a forthcoming paper. By design, VNSIM is not limited to VLBI networks, but is in general applicable to connected-element interferometers, such as VLA. Considering the usability for non-VLBI astronomers without much interferometric knowledge, VNSIM has been designed to be more friendly in user interfaces and more convenient in database management. All kinds of parameters can also be user specified to investigate more simulation scenarios. The comparison of the simulation results from VNSIM with other tools verified the consistency between them. The current version of VNSIM provides functionality matching the astronomers' basic requirements for scheduling and evaluating ground-based interferometric observations. More sophisticated functions are under development. In the future version, we aim to provide more precise space VLBI simulation with realistic constraints and complete scheduling plans.



\acknowledgments

The simulations were performed on the Data Processor Prototype of the China SKA Regional Center hosted by Shanghai Astronomical Observatory with funding support from the Ministry of Science and Technology of China (grant No. SQ2018YFA040022) and the Chinese Academy of Sciences (CAS, grant No. 114231KYSB20170003). T.A. thanks the youth innovation promotion association of the CAS. The authors are grateful to Zsolt Paragi for his help about VLBI network parameter evaluation, and Junyi Wang and Sandor Frey for their constructive comments on the simulations of space VLBI.


\end{document}